\documentstyle[epsfig,graphicx,12pt,epsf]{article}

\newcommand{\nn}{\nonumber}
\newcommand{\NP}[1]{ Nucl.\ Phys.\ {\bf #1}}

\newcommand{\PL}[1]{ Phys.\ Lett.\ {\bf #1}}

\newcommand{\PR}[1]{Phys.\ Rev.\ {\bf #1}}
\newcommand{\PRL}[1]{ Phys.\ Rev.\ Lett.\ {\bf #1}}

\newcommand{\bi}{\bibitem}
\newcommand{\vs}{\vspace{-0.20cm}}
\newcommand{\be}{\begin{equation}}
\newcommand{\ee}{\end{equation}}
\newcommand{\ba}{\begin{eqnarray}}
\newcommand{\ea}{\end{eqnarray}}

\begin{document}

\begin{center}
{\Large{\bf Pion and kaon role in $\tau$ decays, ($g_{\mu}-2$),
 the running of $\alpha_{QED}$ and the muonium hyperfine splitting }}

\vspace{0.3cm}

\end{center}


\begin{center}
{\large{J. E. Palomar}}
\end{center}

\begin{center}

{\small{\it Departamento de F\'{\i}sica Te\'orica and IFIC, \\
Centro Mixto Universidad de Valencia-CSIC, \\
Ap. Correos 22085, E-46071 Valencia, Spain}}


\end{center}

\vspace{0.0003cm}

\begin{abstract}
We make use of recent accurate results obtained for the pion and kaon vector
 form
factors within a chiral unitary approach in order to
 calculate the decay widths of the $\tau$ lepton to these mesons and
 also to evaluate the contribution of this two mesons to the anomalous
 magnetic moment of the muon, the running of the fine structure constant and the
 muonium hyperfine splitting. 
\end{abstract}

\section{Introduction}

In this paper we apply the pion and kaon vector form factors of reference
 \cite{ffoop} to calculate the tau decay 
 to these mesons and also to 
 study their contributions to the hadronic part of the
  anomalous magnetic moment of the muon, the running of the effective 
  structure constant and the muonium hyperfine splitting. All these applications
  should be viewed as a complement of ref.~\cite{ffoop} where a coupled-channel
  non-perturbative chiral approach was used to calculate the final state
  interactions in the pion and kaon vector form factors.
  
  The decay of the $\tau$ lepton into a tau neutrino plus hadrons provides a
unique framework to study low energy QCD. The mass of this lepton, about 1.8
GeV, allows the application of perturbative QCD to study inclusive decays, and
in fact offers the possibility to measure the strong coupling constant
$\alpha_{s}(\mu)$ at the low scale $\mu=m_{\tau}$ \cite{braa}. However,
the calculation of exclusive semileptonic decays is not possible nowadays within
perturbative QCD. One can then try to apply chiral perturbation theory \cite{GyL}, but its
range of applicability is far below $m_{\tau}$. This means that it is possible
 to calculate differential decay rates at low energies, but predictions of
integrated rates are not possible within standard $\chi PT$. A study of the
differential decay rates of $\tau$ decaying to two and three pions was done in
ref.~\cite{taucol} using standard $\chi PT$. The more problematic
resonance region in the
two meson decay mode has been theoretically studied using vector meson dominance
\cite{Finkemeier:1995sr,Kuhn:1990ad} or unitary techniques 
\cite{Beldjoudi:1994hi}. Here we will try to apply the
form factors obtained in ref.~\cite{ffoop} to calculate the decay rates of the
 decays $\tau^{-}\rightarrow \pi^{-}\pi^{0}\nu_{\tau}$ and 
 $\tau^{-} \rightarrow K^{-}K^{0}\nu_{\tau}$. The fact that in these decay modes
 only the $I=1$ current is involved is useful since from the two pion decay mode
 data one can extract the pion form factor free from the $I=0$ contamination due
 to the $\rho$-$\omega$ mixing.
 
 The other observables to be studied in this paper are the anomalous magnetic
 moment of the muon $a_\mu \equiv (g_{\mu}-2)/2$, the running of the effective
  coupling constant of QED and the muonium hyperfine splitting. The anomalous 
  magnetic moment of the muon provides one of the most precise
 tests of the Standard Model (SM). The evaluation of $a_{\mu}$ within the SM has
  three sources:
 QED, electroweak (EW) and hadronic. Recently, the E821 BNL experiment obtained an 
 experimental value
 which presented a 2.6 $\sigma$ deviation from the SM calculation. This
 discrepancy can be interpreted as
 a signal of new physics and it has  
 stimulated a lot of publications in the last year. However, recent
 reevaluations \cite{Knecht:2001qf,Knecht:2001qg,Hayakawa:2001bb} of the 
 pseudoscalar pole contribution to $a_\mu$, correcting a mistake in its sign,
 reduce the discrepancy to 1.6 $\sigma$. The different SM contributions and also
 some possible sources of discrepancy based on supersymmetric loop effects are
 scrutinized in \cite{Czarnecki:2001pv}. The hadronic contribution 
 provides the main source of error in the confrontation between theory and
 experiment. In fact, a first principle QCD calculation is not available, and
 one has to resort to the use of dispersion relations relating the anomaly to
 the cross section for production of hadrons in $e^{+}e^{-}$ annihilations. 
 There is a large amount of such kind of analysis (see \cite{Narison:2001jt} and
 references therein). They find that the hadronic contribution is dominated by
the low energy regime, being the $\rho$ contribution about a 72$\%$ of the total
hadronic effect (the weight of the different energy regions can be seen for
instance in figure 6 of ref.~\cite{Jegerlehner:2001wq}). Our aim here is to
study the pion and kaon contribution to the hadronic part of $a_{\mu}$. The pion
contribution to this magnitude and to the running of the effective structure
coupling is studied also in ref.~\cite{pedanies} making use of elastic unitarity
in the two pion channel.

Another magnitude to study in this paper is the effect of pions and kaons in 
 the running 
of the effective coupling of QED. A good knowledge of this parameter is crucial
in precision physics since it is one of the basic input parameters of the SM.
The vacuum polarization effects are responsible for a partial screening of the
fine structure constant in the Thomson limit, while at higher energies the
strength of the electromagnetic interaction increases. Vacuum polarization is
dominated by the QED contribution which is very well known. Again, the problem is
that the calculation of the low energy contributions of the loop of two quarks
cannot be performed within perturbative QCD, and one has to resort again 
to dispersion relations and the analysis of $e^{+}e^{-}$ data. Recently the different 
hadronic contributions have been reevaluated with such techniques
\cite{Jegerlehner:2001wq,Narison:2001xj}.

Finally we also study in section 5 the effect of pions and kaons in the muonium hyperfine
splitting. The hyperfine structure of two-body systems has attracted the
interest of both experimentalists and theoreticians since it provides precise
tests of bound-state QED and accurate determinations of fundamental constants
like the muon to electron mass ratio. Theoretically the case of muonium is
interesting since it has not the problem of proton structure present in other
two-body atomic systems like the hydrogen, and it has been measured very
precisely despite the short muon lifetime. As in the former observables, the
hadronic contributions are the main theoretical problem and are evaluated
through an analysis of $e^{+}e^{-}$ data using dispersion relations. Recent
analysis can be found in \cite{Narison:2001xj,Czarnecki:2001yx}. In the first
reference the muon to electron mass ratio is also determined in very good
agreement with the experimental value.

\section{Pion and kaon vector form factors}

In this section we briefly review the calculation of the pion and kaon vector form
factors done in ref. \cite{ffoop}. These form factors are the necessary 
information to estimate the contributions of the aforementioned mesons to the
muon anomalous magnetic moment, the running of the effective fine
structure constant and the muonium hyperfine splitting, and to calculate the branching ratios of the $\tau$ decay to
such mesons. The approach of reference \cite{ffoop} calculates the final state
interaction corrections to the tree level
 amplitudes, calculated from lowest order $\chi PT$
\cite{GyL} and from the inclusion of explicit resonance fields \cite{EGPdR},
while matching with the $\chi PT$ vector form factors calculated at
next-to-leading order \cite{GyL}. In \cite{ffoop} it is shown that starting from 
the unitarity
of the $S$-matrix for definite isospin and using matrix notation (since the pion and kaon channels
couple in $I=1$) it is possible to write the following
equation for the form factor $F(s)$:

\begin{center}
\be
\textrm{Im} F^{I}(s)=\tilde{Q}(s)^{-1}\cdot T^{I}(s)\cdot
\frac{Q(s)}{8\pi\sqrt{s}}\cdot\tilde{Q}(s)\cdot F^{I*}(s)
\label{unitff}
\ee
\end{center}

\noindent where $Q(s)_{ij}=\sqrt{s/4-m_{i}^{2}}\theta(s-4m_{i}^{2})
\delta_{ij}$, $\tilde{Q}(s)_{ij}=\sqrt{s/4-m_{i}^{2}}\delta_{ij}$ and $T(s)$ is
a matrix containing the meson-meson scattering amplitudes. In the $I=1$ channel
we have pions, kaons and the $\rho$ resonance, while in the $I=0$ channel we
have kaons and the $\omega$ and $\phi$ resonances.

If one uses in the
former equation the $T$-matrix expression provided by the $N/D$ method adapted
 to
the chiral framework (see ref. \cite{nd}), it is possible to show that the form
factor must have the form:

\begin{center}
\be
F^{I}(s)=[1+\tilde{Q}(s)^{-1}\cdot K^{I}(s)\cdot \tilde{Q}(s)\cdot 
g^{I}(s)]^{-1}\cdot R^{I}(s)
\label{f2}
\ee
\end{center}

\noindent where $K^{I}(s)$ is a matrix collecting the tree level meson-meson
 scattering 
amplitudes\footnote{Calculated from lowest order $\chi PT$ \cite{GyL} plus 
s-channel vector
 resonance exchange contributions \cite{EGPdR}}, $g^{I}(s)$ is the diagonal 
 matrix given by the loop with two meson propagators (see ref. \cite{nd})
  and $R^{I}(s)$ is a vector whose components are functions
 free of any cut. 
 
 To fix these unknown functions we take a look at the large
 $N_{c}$ limit of eq. (\ref{f2}). 
 In this limit loop physics is suppressed, therefore from eq.~(\ref{f2}) we
 find that   
 $F^{I}_{N_{c}^{leading}}(s)=R_{N_{c}^{leading}}(s)=F^{I}_{t}(s)$, where 
 $F^{I}_{t}(s)$ is the 
 tree level form factor\footnote{The tree level form factors are also evaluated
  using the lowest order $\chi PT$ Lagrangian \cite{GyL} plus the chiral
   resonance Lagrangian \cite{EGPdR}}. Once the $N_{c}$-leading part of
   $R^{I}(s)$ is known only its subleading part remains to be fixed. This 
   part should be a polynomial since it has no cuts and the poles coming from
   the resonances are already included in the leading part. If we further
   require that the form factors of eq.~(\ref{f2}) vanish in the limit
   $s\rightarrow \infty$ we find that the polynomials are in fact constants. 
 
 At this moment there are several unknown parameters: the $R^{I}_{subleading}$,
 the $d^{I}_{i}$ parameters appearing in the $g^{I}(s)$ matrix (see
 refs.~\cite{ffoop,nd}) and also the bare masses of the resonances. The bare
 masses of the resonances can be fixed by the requirements that the moduli of
 the scattering amplitudes have a maximum at the energy
 $\sqrt{s}=M^{physical}_{resonance}$, and the rest of the parameters are fixed
 by matching our results with the ones of $\chi PT$ vector form factors at the
 one loop level in the chiral counting applied to both schemes. Finally, the isospin violation effects can be
 introduced via the $\rho$-$\omega$ mixing.
 
 This method provides a good description of the pion and kaon form factors up to
 1.2 GeV and compares very well with the two-loop $\chi PT$ prediction of
 \cite{taucol,ulf2}. For higher energies the effect of other channels like $\omega
 \pi$, $4\pi$, etc., and other resonances ($\rho^{'}$,
 $\rho^{''}$, $\omega^{'}$, $\phi^{'}$, etc.)  becomes
 relevant. The inclusion of these resonances is straightforward in the
 formalism, but since their masses and couplings are not well known they lead to
 a big number of free parameters. 
 
 It is worth saying that, although we use exactly the same expressions for the
 form factors as in \cite{ffoop}, in the calculations here we use different
 values for some parameters. The value of $f=87.4$ MeV used in \cite{ffoop} was calculated
 from eq.~(7.14) of the last reference in \cite{GyL}, relating $f_{\pi}$ and $f$. Here we have recalculated
 that relation starting from eq.~(7.13) of the former reference and using the
 experimental value of $\langle r^{2} \rangle_{S}^{\pi}$, thus obtaining
 $f=86.6\pm 0.5$ MeV. For $F_V$ we use the value $F_V=153\pm 4$ MeV from the $\rho$
 decay to $e^{+}e^{-}$. Finally, in \cite{ffoop} we used $G_V=53$ MeV. This 
value was taken from an estimation of the chiral corrections in the vector form
 factor by means of vector meson dominance plus one loop $\chi PT$ corrections.
 Since our approach includes the former evaluation but also higher orders in the
 $\chi PT$ expansion we have fitted this value to better reproduce the
 experimental data of the pion form factor, finding a value of 
 $G_V=55.52\pm 0.12$ MeV. The errors in the results quoted in the following
 sections are obtained by summing in quadrature the errors from the
 uncertainties in each  of the fundamental parameters $f$, $F_V$, $G_V$.

\section{$\tau$ decay to $\pi^{-}\pi^{0}\nu_{\tau}$ and to
$K^{-}K^{0}\nu_{\tau}$}

In this section we apply the form factors calculated on the previous section to
the calculation of the branching ratios of the decays  $\tau^{-}\rightarrow
\pi^{-}\pi^{0}\nu_{\tau}$ and $\tau^{-}\rightarrow K^{-}K^{0}\nu_{\tau}$. Let us
start with $\tau^{-}\rightarrow \pi^{-}\pi^{0}\nu_{\tau}$. Its amplitude can be
calculated using standard techniques and one has:

\begin{center}
\ba
\Gamma(\tau^{-}\rightarrow
\pi^{-}\pi^{0}\nu_{\tau}) & = & \frac{G_{F}^{2}\cos^{2}\theta_{c}}{384\pi^{3}}
m_{\tau}^{3}\int_{4m_{\pi}^{2}}^{m_{\tau}^{2}}dp^{2} \left(1-
\frac{p^{2}}{m_{\tau}^{2}}\right)^{2}\left(1+
\frac{2p^{2}}{m_{\tau}^{2}}\right)^{2}\nn \\ & &\left(1-\frac{4m_{\pi}^{2}}{p^{2}}
\right)^{3/2}|F_{\pi}(p^{2})|^{2}
\ea
\end{center}

We expect to obtain a rather good result for the total width in spite of not 
having a very good description of the pion form factor up to $s=m_{\tau}^{2}$
since the neutrino $\nu_{\tau}$ carries in average a sizeable fraction of the
energy and hence there is a smaller energy left for the $\pi \pi$ system, and
furthermore the form factor is dominated by  the $\rho$ resonance which is well
described in our model. For energies close to the $\tau$ mass the form factor is
very small and the contribution of this region is not so relevant. In
ref.~\cite{ffoop} our pion form factor is compared to the experimental one obtained from an
analysis of tau decay data. The result we get for the integrate width is:

\begin{center}
\be
\Gamma(\tau^{-}\rightarrow \pi^{-}\pi^{0}\nu_{\tau})=(5.5 \pm 0.3)\cdot 10^{-10}\textrm{ MeV \
\ \ \ \ \ }\textrm{BR}=0.244\pm 0.012
\ee
\end{center}

\noindent to be compared to the PDG value $\Gamma(\tau^{-}\rightarrow \pi^{-}
\pi^{0}\nu_{\tau})/\Gamma_{\tau}^{tot}=(25.40\pm0.14)\%$ \cite{pdg}. The
agreement between this value and our calculation is remarkable.

Now we can study the decay of the $\tau$ lepton to a pair of kaons. In this case
we do not expect a priori such a good result. On one hand the threshold of
production of kaons, around 1 GeV, where the $I=1$ pion vector form factor
calculated in ref.~\cite{ffoop} begins to deviate from data. On the other hand,
the $\phi$ resonance, which dominates the kaon form factor for such energies,
does not show up here because only the $I=1$ part contributes. As pointed out in
the previous section, our approach can be improved with the inclusion of more
massive vector resonances, such as the $\rho'(1450)$, $\rho''$ and so on
\cite{pdg}. However, one has to face then the problem that for these energies
channels like $4\pi$, $\omega \pi$, etc... are no longer negligible and indeed
dominate the width of the latter resonances. As a result we should include more
free parameters to describe these resonances, such as couplings, widths, etc...,
which makes us decide not to include these resonances and see what comes
out with our previous calculation. The width of the process is then given by:

\begin{center}
\ba
\Gamma(\tau^{-}\rightarrow
K^{-}K^{0}\nu_{\tau}) & = & \frac{G_{F}^{2}\cos^{2}\theta_{c}}{768\pi^{3}}
m_{\tau}^{3}\int_{4m_{K}^{2}}^{m_{\tau}^{2}}ds \left(1-
\frac{s}{m_{\tau}^{2}}\right)^{2}\left(1+
\frac{2s}{m_{\tau}^{2}}\right)^{2}\nn \\ & &\left(1-\frac{4m_{K}^{2}}{s}
\right)^{3/2}|F_{K}(s)|^{2}
\ea
\end{center}

We get the following result:

\begin{center}
\be
\Gamma(\tau^{-}\rightarrow K^{-}K^{0}\nu_{\tau})=(2.8 \pm 0.4)\cdot 10^{-12}\textrm{ MeV \
\ \ \ }\textrm{BR}=(1.25\pm 0.13)\cdot 10^{-3}
\ee
\end{center}

\noindent to be compared to the PDG value for the branching ratio of $(1.55\pm 0.17)\cdot 10^{-3}$
\cite{pdg}. Our result is a bit lower than the experimental one but still in
agreement within errors.

\begin{figure}[ht]
\centerline{\includegraphics[width=0.7\textwidth]{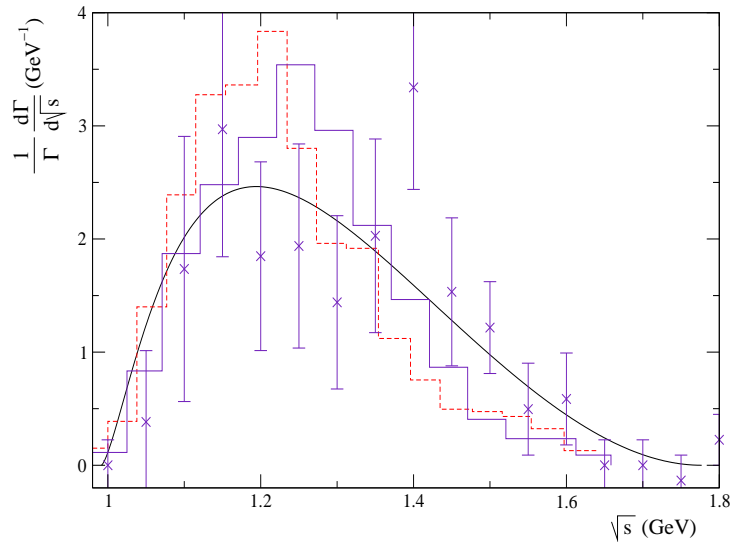}}
\caption{{\small $K\bar{K}$ invariant mass distribution for the decay 
$\tau^{-}\rightarrow K^{-}K^{0}\nu_{\tau}$. Solid line: our calculation; dashed
histogram from ref.~\cite{barate2}. Data and solid histogram from~\cite{coan}. }} 
\label{tauk}
\end{figure}

We have also studied the invariant mass distribution of the $K\bar{K}$ system. The results are plotted
in fig~\ref{tauk}, from were we can see that although the integrated width seems
 a
bit low, our mass distribution is compatible with present data in shape and
strength.

\section{Anomalous magnetic moment of the muon}

The anomalous magnetic moment of the muon can be calculated and measured with
high accuracy, providing an extraordinary test of the electroweak theory. Any
residual difference between the sum of the Standard Model (SM) contributions and
the experimental value $a_{\mu}^{exp}$ will be indicative of new physics. This
observable is being widely studied nowadays and many works have appeared
recently because the last experiment performed at Brookhaven \cite{brookexp} has
obtained a value higher than most of the SM predictions:

\begin{center}
\be
a_{\mu}^{exp}=11659202(14)(6)\cdot 10^{-10}
\label{amuexp}
\ee
\end{center}

The theoretical calculation contains terms of different origin. We may write
$a_{\mu}^{th}$ as a sum of the QED, weak and hadronic contributions. It is
important to assess accurately these contributions in order to see if a new
 contribution from extensions of the SM is needed.

\begin{figure}[ht]
\centerline{\includegraphics[width=0.3\textwidth]{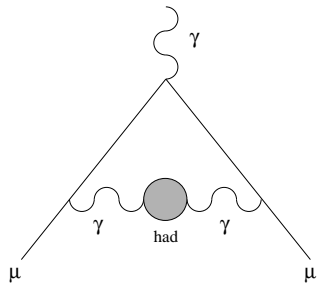}}
\caption{{\small Leading hadronic vacuum polarization contribution to $a_{\mu}$.
The "blob" represents the irreducible photon self-energy.}  }
\label{amups}
\end{figure}

Our aim here is to study the pion, charged kaon and neutral kaon contributions
to the hadronic part of $a_{\mu}$. The leading hadronic contribution to $a_{\mu}$
is due to the photon vacuum polarization insertion into the diagram of the
electromagnetic vertex of a muon, shown in fig~\ref{amups}. This contribution
can be calculated in terms of the experimental cross section
$\sigma_{had}(e^{+}e^{-}\rightarrow hadrons)$ by using dispersion relations
\cite{13j,14j,15j}:

\begin{center}
\ba
a_{\mu}^{had}=\left( \frac{\alpha(0)m_{\mu}}{3\pi}\right)^{2}
\int_{4m_{\pi}^{2}}^{\infty} \frac{ds}{s^{2}}R(s)\hat{K}(s)\textrm{, \ \ \ with}\nn \\
R(s)=\frac{3s}{4\pi \alpha^2(s)}\sigma(e^{+}e^{-} \rightarrow hadrons)
\label{amuhad}
\ea
\end{center}4

\noindent where the $\hat{K}(s)$ function is \cite{jeg}:

\begin{center}
\ba
\hat{K}(s)&=&\frac{3s}{m_{\mu}^{2}}K(s)\textrm{ \ \ \ where}\nn \\
K(s)&=&\frac{x^{2}}{2}(2-x^{2})+\frac{(1+x^{2})(1+x)^{2}}{x^{2}}\left(\ln(1+x)
-x+\frac{x^{2}}{2}\right) +\nn \\
& &+\frac{1+x}{1-x}x^{2}\ln(x)  \textrm{;\ \ \ \ \ \ \ with}\nn \\
& & x\equiv \frac{1-\beta_{\mu}(s)}{1+\beta_{\mu}(s)} \textrm{; \ \ \ \ \ \ \ \ \ }
\beta_{\mu}(s)=\sqrt{1-4m_{\mu}^{2}/s}
\label{kdefi}
\ea
\end{center}

Since $\hat{K}(s)$ grows smoothly, the integral is dominated by the low energy
region.

We have now all the ingredients to calculate the pionic and kaonic
contributions. The cross section $\sigma(e^{+}e^{-}\rightarrow \pi^{+}\pi^{-})$
is:

\begin{center}
\be
\sigma(e^{+}e^{-}\rightarrow \pi^{+}\pi^{-})=\frac{\pi
\alpha^{2}\sigma_{\pi}^{3}}{3s}|F_{\pi}(s)|^{2}
\label{sigmapi}
\ee
\end{center}

\noindent where $\sigma_{\pi}=\sqrt{1-\frac{4m_{\pi}^{2}}{s}}$. Finally the
integral we have to calculate is:

\begin{center}
\be
a_{\mu}^{\pi \pi}=\left( \frac{\alpha(0) m_{\mu}}{6\pi}\right)^{2}
\int_{4m_{\pi}^{2}}^{\Lambda^{2}} \frac{ds}{s^{2}}
\sigma_{\pi}^{3}|F_{\pi}(s)|^{2}\hat{K}(s)
\label{api}
\ee
\end{center}

The results we get for different values of the cut off $\Lambda$ are given in
table \ref{pionamu}. They are to 
be compared to results coming from experimental analysis in the
region $0.320\textrm{ GeV}\le \sqrt{s} \le 2.125\textrm{ GeV}$: 
$(500.81\pm 6.03)\times 10^{-10}$ \cite{alemany} and $(510\pm 5.3)\times 10^{-10}$
 \cite{anderson}. The agreement between our prediction and
the experiment is remarkable. We also agree with the theoretical estimation done
in \cite{pedanies}. The recent analysis of reference \cite{Narison:2001jt} gives
for the pion contribution in the region $4m_{\pi}^{2} \le s \le 0.8$ GeV$^{2}$ 
a value
of $a_{\mu}^{\pi \pi}=(479.46\pm 6.07)\times 10^{-10}$, while in the same energy
 interval we find $(490\pm 18)\times 10^{-10}$, in agreement with that reference.   

\begin{table}
\begin{center}
\begin{tabular}{|c|ccccc|}
\hline
$\Lambda$(GeV) & 1.1 & 1.2 & 1.3 & 1.4 & 2.1 \\
\hline
$a_{\mu}^{\pi \pi}\times 10^{10}$& $512\pm 19$ & $515\pm 20$ & $516\pm 20$ &
$517\pm 20$ & $518\pm 20$ \\
\hline
\end{tabular}\\ 
\hspace{0.5cm}
\caption{Values of $a_{\mu}^{\pi \pi}$ obtained for different values of the cut
off.}
\label{pionamu}
\end{center}
\end{table}

The calculation for the kaons is analogous to the former one. We obtain for a
value of the cut-off $\Lambda$=1.2 GeV the values of $a_{\mu}^{K^{+}K^{-}}\times
10^{10}=18.1 \pm 1.0$ and $a_{\mu}^{K^{0}\bar{K}^{0}}\times 10^{10}= 10.7\pm 0.6
$.

The analysis of data done in ref.~\cite{alemany} gives the values
$a_{\mu}^{K^{+}K^{-}}\times 10^{10}=4.30\pm 0.58$ and
$a_{\mu}^{K^{0}\bar{K}^{0}}\times 10^{10}=1.20\pm 0.42$ in the region 
$1.055\textrm{ GeV}\le \sqrt{s} \le 2.055\textrm{ GeV}$ for $K^{+}K^{-}$ and
$1.090\textrm{ GeV}\le \sqrt{s} \le 2.055\textrm{ GeV}$ for $K^{0}\bar{K}^{0}$. We have integrated also
in these intervals of energies, finding $a_{\mu}^{K^{+}K^{-}}\times
10^{10}=3.2\pm 0.3$
and $a_{\mu}^{K^{0}\bar{K}^{0}}\times 10^{10}=0.233\pm 0.013$. Our values are lower but
this is not strange since in this region of energies there are more resonances
and channels that are not considered in our approach, so that our form factors
are not too accurate for energies higher than 1.2 GeV, specially in the kaon
case. In any case the biggest
contribution comes from energies close to the $\phi$ peak, therefore  it is more
interesting to compare our results in the region where the $\phi$ resonance
dominates the form factor. In reference \cite{alemany} the contribution of the
$\phi$ resonance to $a_{\mu}^{had}$ was studied in the region $1.000\textrm{ GeV}
\le \sqrt{s} \le 1.055\textrm{ GeV}$, giving a value
$a_{\mu}^{\phi}\times
10^{10}=39.23\pm 0.94$. This has the contribution not only from
$K^{+}K^{-}$ and $K^{0}\bar{K}^{0}$ pairs but also from other decay channels of
the $\phi$ as the $\pi^{+}\pi^{-}\pi^{0}$, $\rho \pi$, etc. The 
 $K^{+}K^{-}$ branching ratio is $49.2\%$ and the $K^{0}\bar{K}^{0}$ one is 
 $33.8\%$. We have then that the total branching ratio of 
 $\phi\rightarrow K\bar{K}$ is $83\%$. Since these partial widths are also
 related to the form factor, we can make an estimation of the
 contribution of kaons in this range of energies by multiplying the $\phi$
 contribution by 0.83, obtaining in this way a kaon contribution of approx. 
 $32.6\times 10^{-10}$.
 Carrying out the integral with our form factor in this region of energies we
 find $a_{\mu}^{K^{+}K^{-}}\times 10^{10}=16.3\pm 0.8$ and
 $a_{\mu}^{K^{0}\bar{K}^{0}}\times 10^{10}=10.3 \pm 0.5$, which give a total kaon
 contribution of $(26.5\pm 1.5)\times 10^{10}$, close to the former estimation. 
 The fact that we
 do not get very good values in the case of the $\phi$ is not surprising since
 there are long standing problems related to these resonance like the
 $\Gamma(\phi \rightarrow K^{+}K^{-})/\Gamma(\phi \rightarrow K^{0}\bar{K}^{0})$
 problem \cite{bramon}, and in
 the calculation of the widths the kaon vector form factor is also involved. 
  
\section{$\pi$ and $K$ contributions to the running effective
fine structure constant and the muonium hyperfine splitting.}

\begin{figure}[ht]
\centerline{\includegraphics[width=0.3\textwidth]{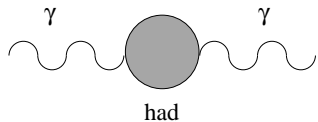}}
\caption{{\small Hadronic contribution to photon vacuum polarization.} } 
\label{vacpol}
\end{figure}

In this section we estimate the pion and kaon contributions to the running of
the fine structure constant and to the muonium hyperfine splitting. The 
effective structure constant at scale $\sqrt{s}$ is given by:

\begin{center}
\be
\alpha(s)=\frac{\alpha}{1-\Delta \alpha(s)}
\label{runa}
\ee
\end{center}

\noindent where $\alpha$ is the fine structure constant and $\Delta \alpha$ is
the photon vacuum polarization contribution. The hadronic contribution to the
photon vacuum polarization corresponds to the "blob" in fig.~\ref{vacpol}.

The low energy contribution to the hadronic part cannot be calculated from
perturbative QCD, but it can be related to the $e^{+}e^{-}$-annihilation data by
using dispersion relations and the optical theorem \cite{30j,39j}, as it is
usually done in the case of $a_{\mu}$. The corresponding expression can be
written in the form of eq.~\ref{amuhad} but replacing $a_{\mu}^{had}$ by $\Delta
\alpha_{had}(M_{Z}^{2})$, and the function $K(s)$ (defined in
eq.~\ref{kdefi}), by

\begin{center}
\be
K_{\alpha}(s)=\frac{\pi}{\alpha}\frac{M^{2}_{Z}}{M^{2}_{Z}-s}
\label{defikalpha}
\ee
\end{center}

Using this equation we get the results shown in table \ref{pionalpha}. This
results are to be compared  
with the result coming from experimental analysis in the
region $0.320\textrm{ GeV}\le \sqrt{s} \le 2.125\textrm{ GeV}$: 
$(34.31\pm 0.38)\times 10^{-4}$ \cite{alemany} . As we can see, 
we have a good agreement with the experiment and also with the calculation done
in ref.~\cite{pedanies}. Also, in the region $4m_{\pi}^{2} \le s \le 0.8$
GeV$^{2}$ we get $\Delta \alpha_{\pi^{+}\pi^{-}}(M_{Z}^{2})=31.7 \pm 1.3$, in
agreement with the recent estimation of \cite{Narison:2001xj} $\Delta
\alpha_{\pi^{+}\pi^{-}}(M_{Z}^{2})=31.45 \pm 0.23$.

\small{\begin{table}
\begin{center}
\begin{tabular}{|c|ccccc|}

\hline
$\Lambda$(GeV) & 1.1 & 1.2 & 1.3 & 1.4 & 2.1  \\
\hline
$\Delta \alpha_{\pi^{+}\pi^{-}}(M_{Z}^{2})\times 10^{4}$& $33.9\pm 1.4$ &
$34.3\pm 1.5$ &
 $34.5\pm 1.5$ & $34.7\pm 1.5$ & $35.1\pm 1.6$ \\
\hline
\end{tabular}\\ 
\hspace{0.5cm}
\label{pionalpha}
\caption{Values of $\Delta \alpha^{\pi^{+}\pi^{-}}(M_{Z}^{2})$ 
 obtained for different values of the cut
off.}
\end{center}
\end{table}}

\normalsize{We} can evaluate in the same way the kaon contribution. We obtain for a cut-off
of $\Lambda=1.2$ GeV the values $\Delta \alpha_{K^{+}K^{-}}(M_{Z}^{2})=(2.43 \pm
0.13)\times 10^{-4}$ and $\Delta \alpha_{K^{0}\bar{K}^{0}}(M_{Z}^{2})=(1.42 \pm
0.07) \times 10^{-4}$.
The experimental data analysis of ref.~\cite{alemany} in the region $1.055
\textrm{ GeV}\le
\sqrt{s}\le 2.055\textrm{ GeV}$ for $K^{+}K^{-}$ and $1.090\textrm{ GeV}\le
 \sqrt{s}\le 2.125\textrm{ GeV}$ for
$K^{0}\bar{K}^{0}$, where the $\phi$ region is excluded, gives  $\Delta \alpha_{K^{+}K^{-}}(M_{Z}^{2})=(0.85\pm 0.10)
\times 10^{-4}$ and $\Delta \alpha_{K^{0}\bar{K}^{0}}(M_{Z}^{2})=(0.23\pm 0.08)
\times 10^{-4}$ respectively. Carrying out the integration in the same energy
regions we obtain $\Delta \alpha_{K^{+}K^{-}}(M_{Z}^{2})=(0.65 \pm 0.07) \times 
10^{-4}$ and
 $\Delta \alpha_{K^{0}\bar{K}^{0}}(M_{Z}^{2})=(0.0392 \pm 0.0023) \times 10^{-4}
 $. As in the $a_{\mu}$ case our results for these intervals of energies are 
 not so good as in the pion case and the reasons are identical.

We can also try to make an estimation of the kaon contribution in the $\phi$
region. Ref.~\cite{alemany} gives a $\phi$ contribution of $\Delta
\alpha_{\phi}(M_{Z}^{2})=(5.18\pm 0.12)\times 10^{-4}$ for 
$1.000\textrm{ GeV}\le \sqrt{s}\le 1.055\textrm{ GeV}$, from where we estimate a
value $ \Delta \alpha^{K\bar{K}}_{\phi}(M_{Z}^{2})\simeq 4.3\times 10^{-4}$. The integral
in this region gives us $\Delta \alpha_{K^{+}K^{-}}(M_{Z}^{2})=(2.15 \pm 0.11) 
\times 10^{-4}$ and $\Delta \alpha_{K^{0}\bar{K}^{0}}(M_{Z}^{2})=(1.35 \pm 0.07)
 \times 10^{-4}$.
Hence the sum, as in the $a_{\mu}$ case, is a bit low.

\begin{figure}[ht]
\centerline{\includegraphics[width=0.7\textwidth]{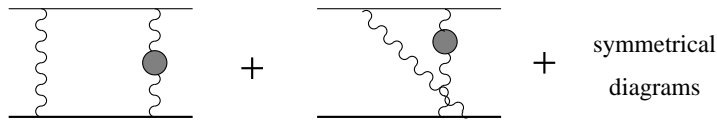}}
\caption{{\small Diagrams accounting for the hadronic contribution to the
muonium hyperfine splitting. }} 
\label{hyperfine}
\end{figure}

Finally, we have studied also the contributions of pions and kaons to the
muonium hyperfine splitting. The hadronic contribution to the hyperfine
splitting of the muonium is given by the diagrams in figure~\ref{hyperfine}. As
in the previous cases, the hadronic blob must be evaluated and in order to do so
one has to resort to dispersion relations.

The ground state hyperfine splitting is given also by eq.~\ref{amuhad}, but
replacing there $a_{\mu}^{had}$ by $\Delta E_{had}$ in te LHS and the function $K(s)$
 (related to $\hat{K}(s)$ as established in
eq.~(\ref{kdefi})) appearing there by (see \cite{Narison:2001xj,Czarnecki:2001yx}):

\begin{center}
\be
K_{split}(s)=\frac{16 \alpha^{4}m_{R}^{3}}{3m_{\mu}^{2}}\left\{ \left(\frac{s}{4
m_{\mu}^{2}}+\frac{3}{2}\right)\textrm{log}\frac{s}{m_{\mu}^{2}}-\frac{1}{2}
-\left(2+\frac{s}{4m_{\mu}^{2}}\right)
\beta_{\mu}(s)\textrm{log}\frac{1+\beta_{\mu}(s)}
{1-\beta_{\mu}(s)}\right\}
\label{khyper}
\ee
\end{center}

\noindent where $m_R$ is the reduced mass and $\beta_{\mu}(s)$ is defined in
eq.~(\ref{kdefi}).

As in the other cases, we have evaluated the pion contribution with different
values of the cut-off, finding the results shown in table \ref{pionhyp}.

\begin{table}
\begin{center}
\begin{tabular}{|c|ccccc|}
\hline
$\Lambda$(GeV) & 1.1 & 1.2 & 1.3 & 1.4 & 2.1  \\
\hline
$\Delta \nu_{\pi^{+}\pi^{-}}\textrm{(Hz)}$& $162\pm 6$ &
$163\pm 6$ &
 $163\pm 6$ & $164\pm 6$ & $164\pm 6$ \\
\hline
\end{tabular}
\hspace{0.5cm}
\label{pionhyp}
\caption{Values of $\Delta \nu_{\pi^{+}\pi^{-}}\textrm{(Hz)}$ 
 obtained for different values of the cut
off.}
\end{center}
\end{table}

As we can see in table \ref{pionhyp}, the pion contribution barely depends on
the value of the cut-off. Here we will compare with the analysis of the
$e^{+}e^{-}$ and $\tau$ decay experimental data done in
ref.~\cite{Narison:2001xj} in the interval $4m_{\pi}^{2} \le s \le 0.8$ 
GeV$^{2}$: $\Delta \nu\textrm{(Hz)}=152.9\pm 1.8$. Our prediction in the same
energy region is $\Delta \nu_{\pi \pi}\textrm{(Hz)}=154\pm 6$, in perfect agreement with
the former estimate. In the case of the kaons we will only give here the result
obtained with a cut-off of $\Lambda=1.2$ GeV since this is the region in which
the kaon form factor is well reproduced. The values obtained are $\Delta
\nu_{K^{+}K^{-}}\textrm{(Hz)}=6.2\pm 0.3$ and $\Delta
\nu_{K^{0}\bar{K}^{0}}\textrm{(Hz)}=3.7\pm 0.2$, and we expect our
results to be also a bit low compared to the data, as in the former cases when studying the kaon
contribution.

\section{Conclusions}

We have applied the formalism developed in \cite{ffoop} to describe the pion and
kaon vector form factors accounting for unitarity in coupled channels to
calculate the contributions of these two mesons to the decay of the $\tau$
lepton and to the hadronic part of the anomalous magnetic moment of the muon,
the running of the fine structure constant and the muonium hyperfine splitting 
at low energies where a calculation within perturbative QCD is not available.
The evaluation of the branching ratios of the decays $\tau\rightarrow
\pi^{-}\pi^{0}\nu_{\tau}$ and $\tau\rightarrow K^{-}K^{0}\nu_{\tau}$ are in good
agreement with the experiment, although the
calculations need a good description of the $I=1$ form factors of the mesons up to 1.7
GeV, while the description of the form factors done in \cite{ffoop} is only
good up to 1.2 GeV, due to the opening of more channels and the presence of more
resonances at these energies that are not taken into account there. The previous
agreement is found 
because the form factors are dominated by the $\rho$ resonance, having small
values at high energies, and thus giving a larger weight to the low energy region
where our description of the form factor is good.

We have also evaluated the pion and kaon contribution (in the low energy region)
to the anomalous magnetic moment of the muon, the running of the QED effective
coupling and the muonium hyperfine splitting. The results obtained for the pion
are good when compared to the available experimental data analyses. However, in
the case of the kaons our values are a bit lower than expected, and this can be
due to the fact that the description of the kaon form factors employed here is
done in the isospin limit, using an averaged value for the kaon mass (we use
the physical values of $m_{K^{+}}$ and $m_{K^{0}}$ for phase space
considerations). It is also worth noting that the kaon form factor is essential
in the $\Gamma(\phi\rightarrow K^{+}K^{-})/ \Gamma(\phi\rightarrow
K^{0}\bar{K}^{0})$ problem which is not yet understood (see \cite{bramon}).

We want also to stress that the rather large errors that we get in our
estimations are mainly due to the value of the $F_V$ parameter appearing in the
chiral resonance Lagrangians~\cite{EGPdR}, which is fixed by the $\rho
\rightarrow e^{+}e^{-}$ decay. More accurate measurements of these quantity will
thus be most interesting. Finally, it is worth saying that 
within the formalism of ref.~\cite{ffoop} the 
inclusion of another octet of vector resonances is straightforward, although it
 leads to a rather large amount of free parameters (new $F_V$ and $G_V$
 parameters, bare masses and initial widths in the resonance propagators, since we do not take
 into account all the relevant channels at that energies), thus loosing the 
  predictive power of the approach.  

\section{Acknowledgements}

Useful discussions and careful reading by E. Oset, J. A. Oller and A. Pich are
most acknowledged. I want to acknowledge financial support from Ministerio de Educaci\'on,
Cultura y Deporte. This work is also supported by DGICYT contract number
BFM2000-1326 and the EU EURODAPHNE network, contract ERBFMRX-CT98-0169.

\end{document}